\input harvmac

\def\>{{\rangle}}
\def\<{{\langle}}

\def\a {{\alpha}}
\def\b {{\beta}}

\def\ul{\underline l}
\def\um{\underline m}
\def\un{\underline n}
\def\ua{\underline a}
\def\ub{\underline b}
\def\uc{\underline c}
\def\ud{\underline d}
\def\up{\underline p}
\def\uq{\underline q}
\def\ur{\underline r}

\Title{\vbox{\hbox{IFT-P.070/97, hep-th/9711055}}}
{\vbox{\centerline{\bf
Duality--Symmetric Eleven--Dimensional Supergravity}
\bigskip\centerline{\bf
and its Coupling to M--Branes}}}
\bigskip\centerline{Igor Bandos$^1$, Nathan Berkovits$^2$ and
Dmitri Sorokin$^1$}
\bigskip\centerline{${~}^1$~NSC, Kharkov Institute of Physics and
Technology}
\centerline{Kharkov, 310108, Ukraine} \centerline{e-mail:
bandos,dsorokin@kipt.kharkov.ua} \vskip .2in
\bigskip\centerline{${}^2$~Instituto
de F\'{\i}sica Te\'orica, Univ. Estadual Paulista}
\centerline{Rua Pamplona 145, S\~ao Paulo, SP 01405-900, BRASIL}
\centerline{e-mail: nberkovi@ift.unesp.br}
\vskip .2in

The standard eleven-dimensional supergravity action depends on
a three-form gauge field and does not allow direct coupling to
five-branes. Using previously developed methods, we construct
a covariant eleven-dimensional supergravity action depending on a
three-form and six-form gauge field in a duality-symmetric manner.
This action is coupled to both the M--theory two-brane and five-brane, 
and corresponding equations of motion are obtained. Consistent coupling 
relates $D=11$ duality properties with self--duality properties of the 
M--5--brane. From this duality--symmetric formulation, one derives an 
action describing coupling of the M--branes to standard $D=11$ 
supergravity.

\Date{November 1997}

\newsec {Introduction}

Eleven-dimensional supergravity has recently returned to popularity
because of M-theory conjectures. The standard D=11 supergravity
action is constructed using a graviton, gravitino and three-form
gauge field $A^{(3)}$ \ref\CJS{E. Cremmer, B. Julia and
J. Scherk, {\sl Phys. Lett.} {\bf B76} (1978) 409.}.
Although it is easy
to couple the standard $D=11$ supergravity multiplet to supergravity
solutions corresponding to ``electric'' 2-branes \ref\tob{
M. J. Duff and K. S. Stelle, {\sl Phys. Lett.} {\bf 253B} (1991) 113.},
it is difficult to couple it to supergravity solutions
corresponding to ``magnetic'' 5-branes \ref\fb{R. G\"uven, {\sl Phys.
Lett.} {\bf 276B} (1992) 49.}. But as the analysis of the effective
worldvolume action of the M--theory five--brane has shown \ref\FIBRA{O.
Aharony, {\sl Nucl. Phys.} {\bf B476} (1996) 470 \semi E. Bergshoeff, M.
de Roo and T. Ort\'\i n, {\sl Phys. Lett.} {\bf 386B} (1996) 85.}
\ref\PSTU{P. Pasti, D. Sorokin and M. Tonin, {\sl Phys. Lett.} {\bf
398B} (1997) 41.} \ref\BLN{I. Bandos, K. Lechner, A. Nurmagambetov, P.
Pasti, D. Sorokin and M.  Tonin, {\sl Phys. Rev. Lett.} {\bf 78} (1997)
4332.} \ref\schwarz{M. Aganagic, J. Park, C. Popescu and J. H. Schwarz,
{\sl Nucl.  Phys.} {\bf B496} (1997) 191.} \ref\st{D. Sorokin and P. K.
Townsend, M--theory superalgebra from the M--5--brane, hep-th/9708003.},
the M--five--brane is a dionic object which carries both an ``electric''
and a ``magnetic'' charge.  Thus, for
coupling the M--5--brane it is desirable to have a $D=11$ supergravity
action which would contain $A^{(3)}$ as well as a six-form gauge
field $A^{(6)}$ whose field strength is the Hodge dual of the
four-form field strength. The importance of the six-form gauge field
in eleven-dimensional supergravity was first recognized in
\ref\NTN {H.
Nicolai, P. K. Townsend and P. van Nieuwenhuizen, {\sl Lett. Nuov. Cim.}
{\bf 30} (1981) 315.} and
\ref\DAF{R. D'Auria and P. Fr\`e, {\sl Nucl. Phys.} {\bf B201} (1982)
101.}, and was later related
in
\ref\MA{P.K. Townsend, {\it p-brane democracy}, hep-th/9507048.}
to central charges in the M-theory superalgebra.

In \ref\ALWIS{S. P. de Alwis, Coupling of branes and normalization of
effective actions in string/M--Theory, hep--th/9705139.}, actions for the
bosonic sector of $D=11$ supergravity with $A^{(3)}$ and $A^{(6)}$ and
their coupling to a membrane and a five--brane have been studied in an
approach where the duality relations are imposed as extra constraints at
the level of equations of motion. This approach seems to be not completely
satisfactory since any modification of these actions (such as
self--coupling, coupling to other fields and sources, and quantum
corrections) would require corresponding consistent modification of the
duality constraints, which can be hard to guess if these constraints are
not yielded by the action. For instance, the five--brane action
produces a highly non--linear self--duality condition for a two--form
worldvolume gauge field \ref\perry
{M. Perry and J. H. Schwarz, {\sl Nucl. Phys.} {\bf 498} (1997) 47
\semi J. H. Schwarz, {\sl Phys. Lett.} {\bf B395} (1997) 191.}
\PSTU ~which reduces to one used in \ALWIS
~only in the linear approximation.

The situation with coupling in $D=11$ supergravity is similar to that
of the four-dimensional Maxwell action which is easily coupled to electric
sources but not magnetic sources \ref\DIR{P. A. M. Dirac, {\sl Proc. R.
Soc.} {\bf A133} (1931) 60, {\sl Phys. Rev.} {\bf 74} (1948) 817 \semi
N. Cabibbo and E. Ferrari, {\sl Nuov. Cim.} {\bf 23} (1962) 1647 \semi
J. Schwinger, {\sl Phys. Rev.} {\bf D3} (1971) 880.}
\ref\ZWAN{D. Zwanziger, Phys. Rev. D3 (1971) 880.}
\ref\des{S.  Deser, A.  Gomberoff, M.  Henneaux and C.  Teitelboim,
Phys.  Lett. B400 (1997) 80, hep-th/9702184.}.
The covariant coupling of Maxwell theory to electric and magnetic sources
has recently been studied \ref\sftu{N. Berkovits, Phys. Lett. B395 (1997)
28, hep-th/9610134.} \ref\med{N. Berkovits and R. Medina,
``Pasti-Sorokin-Tonin Actions in the Presence of Sources'',
hep-th 9704093, to appear in Phys. Rev. D.}
using two different approaches.

The first approach requires an infinite number
of fields and generalizes the McClain-Wu-Yu-Wotzasek action
for two-dimensional chiral bosons \ref\MWYW{
B. McClain, Y.S. Wu, and
F. Yu, Nucl. Phys. B343 (1990) 689,
C. Wotzasek, Phys. Rev. Lett. 66 (1991) 129.}
\ref\MR{I. Martin and A. Restuccia, Phys. Lett. B323 (1994) 311 \semi
F. P.Devecchi  and M. Henneaux,  {\sl Phys. Rev.} {\bf D45} (1996) 1606.}
\ref\sfto{N. Berkovits, Phys. Lett. B388 (1996)
743, hep-th/9607070.}\ref\sftt{ I. Bengtsson and A. Kleppe, {\sl Int. J. Mod.
Phys.} {\bf A12} (1997) 3397.}.
The second approach was developed in \ref\PST
{P.  Pasti, D.  Sorokin and M. Tonin, {\sl Phys. Rev.} {\bf D52} (1995)
4277.}
\ref\PSTO{P.  Pasti, D.  Sorokin and M. Tonin, {\sl Phys. Rev.}
{\bf D55} (1997) 6292.}
and uses a harmonic-like variable (constructed of a scalar field
derivative) to make manifestly covariant duality--symmetric actions
\ref\fj{R.  Floreanini and R.  Jackiw, {\sl Phys. Rev. Lett.} {\bf 59}
(1987) 1873 \semi M.  Henneaux and C. Teitelboim, in Proc. Quantum
mechanics of fundamental systems 2, Santiago, 1987, p. 79; {\sl Phys.
Lett.} {\bf B206} (1988) 650.} \ref\ss{J. H.  Schwarz and A.  Sen, Nucl.
Phys. B411 (1994) 35.}, a dual form of which was first studied by
Zwanziger \ZWAN ~in application to Maxwell theory.  In D=4, both of these
approaches introduce a second vector gauge field whose on-shell field
strength is the dual to the original Maxwell field strength.

In this paper, these two approaches will be used to construct
duality--symmetric actions for $D=11$ supergravity which produce the
duality relation between $A^{(3)}$ and $A^{(6)}$ as a consequence of their
equations of motion.
Then, by generalizing methods developed in \DIR
\des\med, we couple the supergravity action of the second approach to
the membrane and the 5--brane of M--theory. We observe an interesting
phenomenon of intertwining local D=11 and worldvolume symmetries which
are responsible,
respectively, for
duality properties of $D=11$ supergravity  and
self--duality properties of the M--5--brane.
Upon solving for part
of the duality constraints, one can reduce these actions to the standard
Cremmer--Julia--Scherk $D=11$ supergravity and obtain its consistent
coupling to the M--branes, but cannot produce a version with the gauge
field $A^{(6)}$ alone.

Section 2 reviews the standard D=11 supergravity action.
Section 3 describes the McClain-Wu-Yu-Wotzasek form of this action
which then is truncated to a form containing a single
auxiliary scalar field. This version of the theory is discussed in section
4 and, in section 5, it is coupled to the M-branes. Section 6 contains some
concluding remarks and section 7 is an appendix which discusses
supersymmetry transformations of $A^{(6)}$.

\newsec{Six-form gauge fields in D=11 supergravity}

\subsec{ Review of D=11 supergravity}

Our notation and conventions are close to
\ref\duff{M.J. Duff, B.E.W. Nilsson and C.N. Pope,
Phys. Rep. 130 (1986) 1.}. We use the almost plus signature for the
metric and underlined latin letters for the indices of $D=11$ vectors.
Not underlined latin indices will correspond to M-brane worldvolumes.

The standard eleven-dimensional supergravity action is \CJS:
\eqn\stan{ {\cal S}_{CJS}=\int d^{11}x [ {1\over 4}eR(\omega) -{{ie}\over
2}\bar\Psi_{\um}\Gamma^{\um\un\up}D_{\un}[{1\over 2}(\omega+\tilde\omega)]
\Psi_{\up}}
$$-{1\over{4!7!}}
\varepsilon^{\um_1...\um_4\un_1...\un_7}(C^{(7)}+
{~}^*C^{(4)})_{\un_1...\un_7}
(F^{(4)}-{1\over 2}C^{(4)})_{\um_1...\um_4}
$$
$$
-{e\over{2^. 4!}}F^{(4)}_{\um_1...\um_4}F^{(4)\um_1...\um_4}]
+\int {1\over 3}A^{(3)}\wedge F^{(4)}\wedge F^{(4)},
$$
where $e_{\um}^{~\ua}$ is a vielbein describing coupling to
gravity, $e=\det e_{\um}^{~\ua}$
(letters from the beginning of the alphabet
denote flat tangent space indices) and $g_{\um\un}=e_{\um}^{~\ua}e_{\un\ua}$
is a D=11 metric; $F^{(4)}=dA^{(3)}$ is the field strength of the three-form
gauge field,
$\Psi_{\um}^\alpha(x)$ is the gravitino field ($\alpha$=1,...,32),
$\omega_{\um\ua\ub}$ is a spin
connection with torsion and $D_{\um}(\omega)$
is a corresponding covariant derivative
as defined in
\CJS\duff,
$\tilde\omega_{\um\ua\ub}=
\omega_{\um\ua\ub}+{i\over
4}\bar\Psi^{\un}\Gamma_{\um\ua\ub\un\up}\Psi^{\up}$,
$$
C^{(7)}_{\un_1...\un_7}={{e}\over{4^.4!}}
\varepsilon_{\um_1...\um_4\un_1...\un_7}
\bar\Psi_{\up}\Gamma^{\up\um_1...\um_4q}\Psi_{\uq}=-{{21}\over
2}\bar\Psi_{[\un_1}\Gamma_{\un_2...\un_5}\Psi_{\un_7]}, ~~and
$$
$$
C^{(4)}_{\um_1\um_2\um_3\um_4}=3\bar\Psi_{[\um_1}\Gamma_{\um_2\um_3}
\Psi_{\um_4]},$$
where the antisymmetric product $\Gamma^{(p)}$ of $p$
gamma--matrices $dx^{\um} e_{\um}^{~\ua}\Gamma_{\ua}$ satisfies the Hodge
duality condition $$\Gamma^{(p)}=-(-1)^{{{(11-p)(10-p)}\over 2}}
~{}^*\Gamma^{(11-p)}.  $$

This action is invariant under
gauge transformations of $A^{(3)}$ ($\delta A^{(3)}=d\phi^{(2)}$) and
under the supersymmetry transformations:
\eqn\susy{
\delta e_{\um}^{~\ua}=-i\bar\epsilon\Gamma^{\ua}\Psi_{\um},\quad
\delta A^{(3)}_{\um_1\um_2\um_3}={3\over
2}\bar\epsilon\Gamma_{[\um_1\um_2}\Psi_{\um_3]},}
$$
\delta \Psi_{\um} = D_{\um}(\tilde\omega)\epsilon
-{i\over{3!4!}}(\Gamma_{\um\un_1...\un_4}-8\Gamma_{\un_1\un_2\un_3}
g_{\un_4\um})
{\tilde F}^{(4)\un_1...\un_4} \epsilon,
$$
where
\eqn\scf{{\tilde F}^{(4)} = dA^{(3)} -C^{(4)}\equiv F^{(4)}-C^{(4)}}
is a supercovariant field strength in the sense that
its supersymmetry variation does not contain
derivatives of the supersymmetry parameter $\epsilon^\alpha(x)$ \CJS.

It is easy to couple \stan ~to
electric 2-branes by adding the term
$$ \mu_E \int_{{\cal M}_3} dy^{m_1}\wedge dy^{m_2}
\wedge dy^{m_3} A_{m_1 m_2
m_3}^{(3)} (x(y)) \quad (m=0,1,2), $$
where $y$ parametrizes the three-dimensional worldvolume ${\cal M}_3$
spanned by the 2-brane, $A{(x(y))}$ is the pullback onto ${\cal M}_3$
of the $D=11$ gauge field form and $\mu_E$ is the electric
charge (or membrane tension).

However, as mentioned in the
introduction, it is not straightforward
to couple to 5-branes using the fields in \stan since
minimal coupling of the M-5-brane is described by the term  \FIBRA \PSTU
$$ \mu_M \int_{{\cal M}_6} dy^{m_1}\wedge ...\wedge dy^{m_6}
A_{m_1...m_6}^{(6)}(x(y)) ~~~~~(m=0,1...,5)
$$
where $y$ now parametrizes the
six-dimensional worldvolume ${\cal M}_6$ spanned by the 5-brane,
$A^{(6)}(x(y))$ is the pullback onto ${\cal M}_6$ of a $D=11$ six-form gauge
field whose field strength is dual to $F^{(4)}$, and $\mu_M$ is the
magnetic charge (or five--brane tension).

\subsec{Six-form gauge field}

So one needs to find an action for D=11 supergravity containing
a six-form gauge field whose field strength is dual to $F^{(4)}$.
The duality relation between $A^{(6)}$ and $A^{(3)}$
must be slightly more complicated than $F^{(7)}={}^* F^{(4)}$ since
the equation of motion for $A^{(3)}$ which follows from \stan ~is not $d^*
F^{(4)}=0$, but is instead
$$d^* F^{(4)} = -F^{(4)} \wedge F^{(4)} +d
C^{(7)} + d^* C^{(4)}.$$
Since the equations of motion for $A^{(3)}$
should imply Bianchi identities for the dual field-strength, the
appropriate duality condition for the field strength of $A^{(6)}$ is
\eqn\das{dA^{(6)} = {}^* F^{(4)} + A^{(3)}\wedge F^{(4)} -C^{(7)} -{}^*
C^{(4)}.}

As shown in the appendix,
$A^{(6)}$ must transform as
\eqn\sus{\delta
A^{(6)}_{\um_1...\um_6}=3\bar\epsilon\Gamma_{[\um_1...\um_5}\Psi_{\um_6]}
+\delta A^{(3)}_{[\um_1\um_2\um_3}A^{(3)}_{\um_4\um_5\um_6]}}
in order for
\das ~to be preserved on-shell by spacetime-supersymmetry transformations.

It will be convenient to define in addition to \scf ~the supercovariant
field strength
\eqn\scs{\tilde F^{(7)} = dA^{(6)} -A^{(3)}\wedge dA^{(3)}
+ C^{(7)} = F^{(7)}+C^{(7)}}
where $ F^{(7)}\equiv
 dA^{(6)} -A^{(3)}\wedge dA^{(3)}$.
Then the duality relation \das ~takes the
form
\eqn\dr{\tilde F^{(7)}= {}^* \tilde F^{(4)}.}
For later
consideration, it is also convenient to introduce duality related
``generalized" field strengths
\eqn\four{\tilde{\cal F}^{(7)}\equiv
\tilde F^{(7)}-{}^*\tilde F^{(4)}; \qquad\tilde {\cal F}^{(4)}\equiv
\tilde F^{(4)}+{}^*\tilde F^{(7)}; }
$$
\tilde {\cal F}^{(7)}=-{~}^*\tilde {\cal F}^{(4)}, \qquad
\tilde{\cal F}^{(4)}={~}^*\tilde{\cal F}^{(7)},$$
which are zero when the duality relation \dr ~is satisfied.

An action which yields \dr ~as an equation of motion can be constructed
using two methods. The first method will be discussed in
the following section, and the second method will be described in
section 4.

\newsec{Duality-symmetric action with an infinite number of fields}

One obvious way to obtain \dr ~as an equation of motion is to add the
term
$$\int d^{11} x ~L_0^{(4)} (\tilde F^{(4)} +{}^*\tilde F^{(7)})=
\int d^{11} x ~L_0^{(4)}\tilde{\cal F}^{(4)}
$$
to the action of \stan ~where $L_0^{(4)}$ is an unconstrained four-form
which acts as a Lagrange multiplier. However, the equation of motion
from varying $A^{(6)}$ will imply that $L_0^{(4)}$ describes
a propagating field. To eliminate this undesired propagating field,
one needs
to add to
the action the infinite sum
$\int d^{11} x \sum_{I=0}^\infty L_I^{(4)} L_{I+1}^{(4)}$ where
$L_I^{(4)}$ for $I=0$ to $\infty$ is an infinite set of four--form
auxiliary fields. So the complete action is
\eqn\inf{{\cal S}_{CJS} + \int
d^{11} x ~(L_0^{(4)} \tilde{\cal F}^{(4)} + \sum_{I=0}^\infty
L_I^{(4)} L_{I+1}^{(4)}).}

As discussed in references \sfto,\sftt ~and
\sftu, variation of $L^{(4)}_{I+1}$ implies that
$L^{(4)}_I=0$ for
each $I$, assuming that only a finite number of $L^{(4)}_I$'s are non-zero.
In other words, the only solution to the equations of motion containing
a finite number of non-zero fields is when the original $D=11$
supergravity fields are on-shell and when $L^{(4)}_I=0$ for all $I$.
 The condition that only a finite
number of fields are non-vanishing
can be understood as a discretized version of
the asymptotic boundary condition $L^{(4)} (x)=0$ as $x\to\infty$.
\ref\hull{N. Berkovits and C. Hull, to appear \semi
N. Nekrassov, private communication.}

The supersymmetry transformations which leave \inf ~invariant are
easily found to be\foot
{We would like to thank Hitoshi Nishino for noticing a problem with
the original version of these supersymmetry transformations.}
\eqn\susyo{\delta L_{2I}^{(4)}=0,}
$$\delta L_{2I+1\, \up_1 ... \up_4}
 = - M_{\um_1 ... \um_4 \alpha}^{\uq} N^\alpha_{\uq \up_1 ... \up_4}
L_{2I}^{\um_1 ...
 \um_4} + M_{\up_1 ... \up_4 \alpha}^{\uq} N^\alpha_{\uq \um_1 ... \um_4}
L_{2I+2}^{\um_1
... \um_4} ,$$
$$\delta e_{\um}^{~\ua}=-i\bar\epsilon\Gamma^{\ua}\Psi_{\um},\quad
\delta A^{(3)}_{\um_1\um_2\um_3}={3\over
2}\bar\epsilon\Gamma_{[\um_1\um_2}\Psi_{\um_3]},$$
$$ \delta
A^{(6)}_{\um_1...\um_6}=3\bar\epsilon\Gamma_{[\um_1...\um_5}\Psi_{\um_6]}
+\delta A^{(3)}_{[\um_1\um_2\um_3}A^{(3)}_{\um_4\um_5\um_6]},
$$
$$ \delta
\Psi_{\um} = D_{\um}(\tilde\omega)\epsilon
-{i\over{3!4!}}(\Gamma_{\um\un_1...\un_4}-
8\Gamma_{\un_1\un_2\un_3}g_{\un_4\um})
({\tilde F}^{(4)\un_1...\un_4} -2 L_0^{(4)\un_1...\un_4})\epsilon, $$
where
$$ M_{\um_1 ... \um_4 \alpha}^{\uq} =
d \tilde{\cal F}_{\um_1 ...  \um_4} /d\psi_{\uq}^\alpha $$
and $$
N^{\alpha}_{\uq \up_1 ... \up_4}= d (\delta
\psi_{\uq}^\alpha) / d L_{0}^{\up_1 ... \up_4}.$$
Note that the supersymmetry
transformation of $L^{(4)}_I$
vanishes in the linearized approximation and is ``local'' in the
sense that it only depends on $L^{(4)}_J$ for $I-1\leq J\leq I+1$.
These properties are also true for the infinite set of Ramond-Ramond fields
in the superstring field theory action of \sfto ~for the Type II
superstring.

Although this method for introducing six-form gauge fields into the
$D=11$ supergravity action appears somewhat trivial,
it is interesting to note
that closed superstring field theory uses precisely this method
for describing Ramond-Ramond fields and their coupling to
electric and magnetic D-branes. Since the Type IIA superstring
is conjectured to come from dimensional reduction of some
eleven-dimensional M-theory containing supergravity, perhaps
the closed superstring field theory action of \sfto ~comes from
dimensional reduction of some eleven-dimensional action similar to \inf.
Note, however, that this dimensional reduction can not be
straightforward since the three-form gauge field of \stan ~reduces
to both Ramond-Ramond and NS-NS gauge fields in ten dimensions.

We now turn to the second covariant approach to the description of
duality--symmetric fields. As was shown in \PSTO, this formulation can be
considered as a consistent covariant truncation of the infinite series of the
auxiliary fields in \inf ~by putting all $L_I$ with $I>0$ to zero and
choosing $L^{(4)}_{0\un_1...\un_4}={2\over{(\partial
a)^2}}\partial^{\um}a\tilde{\cal
F}^{(4)}_{\um [\un_1\un_2\un_3}\partial_{\un_4]} a$, where $a(x)$ is a
scalar auxiliary field which appears in the model in a nonpolynomial way.
In some sense, what we have done is that we have hidden the infinite
series into this nonpolynomiality. This leads to one of the forms of the
$D=11$ supergravity action to be considered in the next section.

 \newsec{Covariant duality--symmetric formulation with the $a(x)$--field}

\subsec{Actions for the dual gauge fields}
As we have seen in the previous section, to construct a covariant D=11
supergravity action with both $A^{(3)}$ and $A^{(6)}$ fields,
one should introduce auxiliary
fields. In this section, we shall apply the covariant approach of
\PST\PSTO where covariance is gained by the use of the single auxiliary
scalar field $a(x)$.  It has been proven
convenient to introduce a time--like ``harmonic'' unit vector
\eqn\five
{v_{\um}(x)\equiv
{{\partial_{\um}a(x)}\over{\sqrt{-(\partial^{\ul}a\partial_{\ul}a)^2}}},
\qquad v^{\um}v_{\um}=-1, }
by means of which $a(x)$ enters the action (though we
could just as well use a space--like vector
${{\partial_{\um}a(x)}/{\sqrt{(\partial^{\ul}a\partial_{\ul}a)^2}}}$).

For simplicity, in
this and the next subsection we put the gravitino field to zero. This is
reflected in the absence of `tilde' over the field strengths \scf, \scs
~and \four, which now do not contain $C^{(4)}$ and $C^{(7)}$.

We present three
equivalent forms of the action for the dual $A^{(3)}$ and $A^{(6)}$
field, which are obtained from each other by reordering corresponding
terms, each of them being useful for different purposes:
\eqn\six{ S^1_A= \int
d^{11}x~e[{1\over{2^.3!}}v_{\up}~{}^*F^{(7)\up\um_1\um_2\um_3}
{\cal F}^{(4)}_{\um_1\um_2\um_3\uq}v^{\uq} }
$$
+{1\over{2^.6!}}v_{\up}~{}^*F^{(4)\up\um_1...\um_6}{\cal
F}^{(7)}_{\um_1...\um_6\uq}v^{\uq}
 +\int {1\over 6}
F^{(7)}\wedge F^{(4)}.$$
\eqn\seven{ S^2_A=\int d^{11}x~e[{-1\over{4^.
4!}}F^{(4)}_{\um_1...\um_4}F^{(4)\um_1...\um_4} -{1\over{4^.
7!}}F^{(7)}_{\um_1...\um_7}F^{(7)\um_1...\um_7} }
$$ -{1\over{4^.  3!}}v^{\up}{\cal F}^{(4)}_{\up\um_1\um_2\um_3} {\cal
F}^{(4)\uq\um_1\um_2\um_3}v_{\uq}
-{1\over{4^. 6!}}v^{\up}{\cal F}^{(7)}_{\up\um_1...\um_6}{\cal
F}^{(7)\uq\um_1...\um_6}v_{\uq}]$$
$$+\int {1\over 6}F^{(7)}\wedge F^{(4)}.$$
\eqn\cjs{ S^3_A=-\int d^{11}x~{e\over{2^.4!}}F^{(4)}_{\um_1...\um_4}
F^{(4)\um_1...\um_4}
+\int {1\over 3}A^{(3)}\wedge F^{(4)}\wedge F^{(4)} }
$$ -\int
dx^{11}{e\over{2^. 3!}}v^{\up}{\cal F}^{(4)}_{\up\um_1\um_2\um_3}{\cal
F}^{(4)\uq\um_1\um_2\um_3}v_{\uq}.$$

The form \six ~of the action is the most suitable for deriving
the equations of motion of the gauge fields and $a(x)$, and getting the
duality relations \dr .

The forms \six ~and \seven ~are manifestly duality--symmetric with respect
to $F^{(7)}$ and $F^{(4)}$.
(Note that, because of the definition \scs, $ F^{(7)}\wedge
F^{(4)}$=$- A^{(3)}\wedge F^{(4)}\wedge F^{(4)}$
up to a total
derivative).  It is
convenient to consider a combination of \six ~and \seven ~when checking
local bosonic symmetries of the model which we present in the next
subsection.

Finally, the action in the form \cjs ~is one which we have obtained by
truncating the infinite field action of section 3. Though not
manifestly duality--symmetric, \cjs ~is very close to the $A^{(3)}$
field Lagrangian in the Cremmer--Julia--Scherk action \stan.
It differs from
the latter by the last term, the only place where the field strength
of $A^{(6)}$ is contained.  This form is the most appropriate for verifying
local supersymmetry of the complete D=11 supergravity action in the
formulation considered.

\subsec{Local bosonic symmetries of the action and equations of
motion}
The actions
\six--\cjs ~are general coordinate invariant and possess
the ordinary gauge symmetries
\eqn\eight
{\delta A^{(3)}=d\phi^{(2)}, \qquad \delta A^{(6)}=d\phi^{(6)}
+\phi^{(2)}\wedge F^{(4)},}
as well as additional local symmetries whose presence ensures the duality
relations between the gauge fields \fj \ss, space--time covariance, and
an auxiliary nature of $a(x)$ \PST \PSTO. The corresponding local
transformations of the fields are
\eqn\nine
{\delta
A^{(3)}=da\wedge\varphi^{(2)}, \qquad \delta
A^{(6)}=da\wedge\varphi^{(6)}+ da\wedge\varphi^{(2)}\wedge A^{(3)};}
\eqn\ten {\delta a=\phi(x),\quad \delta
A^{(3)}=-{\phi\over{\sqrt{-(\partial a)^2}}}i_v{\cal F}^{(4)}, }
$$\delta_\phi A^{(6)}={\phi\over{\sqrt{-(\partial a)^2}}}i_v{\cal F}^{(7)}
+\delta_\phi A^{(3)}\wedge A^{(3)},$$
where
\eqn\ele
{i_v{\cal F}^{(4)}\equiv {1\over{3!}}v^{\up}{\cal
F}^{(4)}_{\up\um_1\um_2\um_3}dx^{\um_1}\wedge dx^{\um_2}\wedge dx^{\um_3},}
$$i_v{\cal F}^{(7)}\equiv {1\over{6!}}v^{\up}{\cal
F}^{(7)}_{\up\um_1...\um_6}dx^{\um_1}\wedge ... \wedge dx^{\um_6}$$
are the 3-- and the 6--form obtained by contracting the 4-- and the
7--form field strength with $v^{\up}$ \five.

For varying the action
with respect to $A^{(3)}$, $A^{(6)}$ and $a(x)$, we should know the
variations of the field strengths which are
\eqn\twel {\delta
F^{(4)}=d(\delta A^{(3)}), \qquad \delta F^{(7)}=d(\delta A^{(6)})
-d(\delta A^{(3)}\wedge A^{(3)}) -2\delta A^{(3)}\wedge F^{(4)}.}
Then a general variation of the action is
$$
\delta S_A=\int
d^{11}x[{1\over{3!7!}}\varepsilon^{\um\un_1\un_2\un_3\um_1...\um_7}
v_{\um}({\cal F}^{(4)}_{\un_1\un_2\un_3\up}v^{\up})\delta
F^{(7)}_{\um_1...\um_7} $$
\eqn\thir{
+{1\over{4!6!}}\varepsilon^{\um\un_1...\un_6\um_1...\um_4} v_{\um}({\cal
F}^{(7)}_{\un_1...\un_6\up}v^{\up})\delta F^{(4)}_{\um_1...\um_4}]
+\delta_{a(x)}S_A,} or in terms of differential forms (and up to a total
derivative)
$$ \delta S_A=-\int [v\wedge i_v{\cal F}^{(4)}\wedge\delta
F^{(7)}-\delta A^{(3)}\wedge d (v\wedge i_v{\cal
F}^{(7)})]+\delta_{a(x)}S_A $$
\eqn\fourt{=\int[(\delta A^{(6)}-\delta
A^{(3)}\wedge A^{(3)}- {{\delta a}\over{\sqrt{-(\partial a)^2}}}i_v{\cal
F}^{(7)})\wedge d(v\wedge i_v{\cal F}^{(4)}) } $$ +(\delta
A^{(3)}+{{\delta a}\over{\sqrt{-(\partial a)^2}}}i_v{\cal F}^{(4)})\wedge
(d(v\wedge i_v{\cal F}^{(7)})+2v\wedge i_v{\cal F}^{(4)}\wedge F^{(4)})].
$$
 From \fourt ~we get the equations of motion
of $A^{(3)}$ and $A^{(6)}$:
\eqn\fif {d(v\wedge i_v{\cal F}^{(4)})=0,
\qquad d(v\wedge i_v{\cal F}^{(7)})+2v\wedge i_v{\cal F}^{(4)}\wedge
F^{(4)}=0.}
As usual in this sort of models \fj \ss \PST \PSTO, these equations
reduce to the duality conditions ${\cal F}^{(4)}=0={\cal F}^{(7)}$
\dr ~(with zero gravitino part) upon
gauge fixing the symmetries under \nine, and the equation of motion of
$a(x)$ is not independent but is a consequence of \fif, which reflects its
auxiliary nature \PST \PSTO.

Substituting variations \eight--\ten ~into \fourt ~one can easily check
that they indeed form local symmetries of the action.

Using the $\varphi^{(2)}$-transformations in \nine,
if we now gauge fix $i_v{\cal F}^{(4)}=0$ (but do not use the second
equation in \fif ~containing $i_v{\cal F}^{(7)}$, which is dynamical),
and substitute this condition into \cjs ~we get the standard
Cremmer--Julia--Scherk action. This explains how duality--symmetric
actions reduce to conventional ones \ss \PST. Note, however, that we cannot
eliminate $A^{(3)}$ and get an action only in terms of $A^{(6)}$, since
$A^{(3)}$ enters the actions \six--\cjs ~directly (i.e. not only
through its field strength as $A^{(6)}$ does). This is why a $D=11$
supergravity action with $A^{(6)}$ alone has not been constructed
\NTN \DAF.

\subsec{Complete duality-symmetric action for D=11 supergravity}
To obtain
the D=11 supergravity action in a duality--symmetric form with the
scalar auxiliary field,
we should replace the part of the Cremmer--Julia--Scherk action containing
$A^{(3)}$ with one of the Lagrangians \six--\cjs ~appropriately
modified due to the presence of the gravitino terms. Note
that the inclusion of gravitino terms in the full action must
respect the local bosonic symmetries \nine ~and \ten. The appropriate action is
$$ S=\int
d^{11}x[{1\over 4}eR(\omega) -{{ie}\over
2}\bar\Psi_{\um}\Gamma^{\um\un\up}D_{\un}{{1\over
2}(\omega+\tilde\omega)}\Psi_{\up} $$
\eqn\sixs{ -{1\over{2^.4!7!}}
\varepsilon^{\um_1...\um_4\un_1...\un_7}(C^{(7)}
+{~}^*C^{(4)})_{\un_1...\un_7}]
(F^{(4)}-{1\over 2}C^{(4)})_{\um_1...\um_4}]+{\tilde S}_A.}
In \sixs, $\tilde S_A$ is one of the duality-symmetric actions \six,
\seven, where $F^{(7)}$ is replaced with \eqn\ninet{
F^{(7)}+C^{(7)}+{~}^*C^{(4)}}
everywhere except in $F^{(7)}\wedge F^{(4)}$.
If $\tilde S_A$ is chosen in the form \cjs, then the coefficient
in front of $(C^{(7)}+{}^*C^{(4)})$ in \sixs ~acquires an additional
factor 2 (i.e. is the same as in \stan).

 From the action \sixs, one gets the duality relation \das ~(or \dr) between
\ninet ~and $F^{(4)}$.

Local supersymmetry transformations under which the action \sixs ~is
invariant are:
$$
\delta a = 0, \qquad \delta
e_{\um}^{~\ua}=-i\bar\epsilon\Gamma^{\ua}\Psi_{\um}, \qquad \delta
A^{(3)}_{\um_1\um_2\um_3}={3\over
2}\bar\epsilon\Gamma_{[\um_1\um_2}\Psi_{\um_3]}, $$
\eqn\twen{ \delta
A^{(6)}_{\um_1...\um_6}=3\bar\epsilon\Gamma_{[\um_1...\um_5}\Psi_{\um_6]}
+\delta A^{(3)}_{[\um_1\um_2\um_3}A^{(3)}_{\um_4\um_5\um_6]}.}

The transformations of $e_m^{~a}$ and $A^{(3)}$ are
the same as in the standard
version while, for the gravitino, we have
\eqn\tweno{
\delta \Psi_{\um} = D_{\um}(\tilde\omega)\epsilon }
$$
-{i\over{3!4!}}(\Gamma_{\um\un_1...\un_4}
-8\Gamma_{\un_1\un_2\un_3}g_{\un_4\um})
({\tilde F}^{(4)\un_1...\un_4}-4v^{[\un_1}{\tilde{\cal
F}}^{(4)\un_2\un_3\un_4]\up}v_{\up})\epsilon.$$

Eq. \tweno ~reduces to the standard supersymmetry transformations
when $i_v{\tilde{\cal F}}^{(4)}$ is put to zero.
Note that
\eqn\twent{
{\tilde F}^{(4)}+v\wedge i_v{\tilde{\cal F}}^{(4)}=
-~{}^*({\tilde F}^{(7)}+v\wedge i_v{\tilde{\cal
F}}^{(7)}).}
Using this duality property one can rewrite \tweno ~in a
duality--symmetric form
$$ \delta \Psi_{\um} = D_{\um}(\tilde\omega)\epsilon $$
$$-{i\over{2^.3!4!}}(\Gamma_{\um\un_1...\un_4}-8\Gamma_{\un_1\un_2\un_3}
g_{\un_4\um})
({\tilde F}^{(4)\un_1...\un_4}-4v^{[\un_1}{\tilde{\cal
F}}^{(4)\un_2\un_3\un_4]\up}v_{\up})\epsilon
$$
\eqn\twenth{
+{i\over{3!7!}}(\Gamma_{\um\un_1...\un_7}-{7\over
2}\Gamma_{\un_1...\un_6}g_{\un_7\um}) ({\tilde F}^{(7)\un_1...\un_7}
+6v^{[\un_1}{\tilde{\cal F}}^{(7)\un_2...\un_7]\up}v_{\up})\epsilon.}

To check that the supergravity action \sixs ~is indeed invariant under the
supersymmetry transformations \twen ~and \tweno, it is convenient to
use the form \cjs ~of the dual gauge field action. Then all
standard terms in the supersymmetry variation of \sixs which do not
contain $\tilde{\cal F}^{(4)}$ vanish, as was proved by Cremmer, Julia and
Scherk \CJS, while the variation terms which contain $\tilde{\cal F}^{(4)}$
have a structure similar to the standard terms with $F^{(4)}$ and, hence,
cancel as well.

We should note that the tensor \twent ~is invariant under the gauge
transformations \eight ~and \nine, but its variation with respect to
\ten ~vanishes only on the mass shell \fif. This indicates that
the supersymmetry variations \tweno ~of the gravitino commute with
the bosonic transformations \ten ~ only up to equations of
motion.

A reason why the supersymmetry transformations for gravitino acquire
additional terms with $\tilde{\cal F}$ is that the auxiliary field $a(x)$
is assumed to be invariant under the supersymmetry transformations
\foot{See \ref\DLT{G. Dall'Agata, K. Lechner and M.
Tonin, Covariant actions for N=1, D=6 supergravity theories with chiral
bosons, hep-th/9710127.} ~for a detailed discussion of the modified
supersymmetry transformations in duality--symmetric models.}.

This
implies that the anticommutator of two supertransformations contains not
only the bosonic translation generator $P_{\um}$ (as in the ordinary case)
but also a generator ${\cal G}$ of the local transformation \ten ~with a
value of the parameter such that it cancels the general coordinate
transformation when acting on $a(x)$, the form of the anticommutator of
supercharges being
\eqn\al {\{Q_\alpha,Q_\beta\}=(\Gamma^{\um})_{\alpha\beta}
(P_{\um}-(\partial_{\um} a) {\cal G}).}
Note that the gravitino is
invariant under \ten ~by definition, and its supersymmetry transformations
(and their commutator with \ten) close only on the mass
shell.  Instead of modifying the supersymmetry transformations for
the gravitino, one might try to find a suitable supersymmetry transformation of
$a(x)$. But because $a(x)$ enters the action in a specific way, it
seems problematic
to find such a transformation.  So we use the modified
supersymmetry transformations for fermions as has always been done in the
models of this kind \ss \PSTO \DLT ~\foot{The only known exception is a
model of supersymmetric chiral bosons in $d=2$ for which a standard
superfield formulation was constructed \ref\PSTLe{P.  Pasti, D.  Sorokin
and M.  Tonin, Space--time symmetries in duality symmetric models.  In
Leuven Notes in Mathematical and Theoretical Physics, (Leuven University
Press) Series B {\bf V6} (1996) pp.  167--176, hep--th/9509053.}.}.

\newsec{Coupling to the M-branes}
The super--p--branes naturally couple to
supergravity fields propagating in curved target {\it superspace} (i.e. to
superfields), and the requirement of local {\it kappa}--symmetry of
the super--p--branes puts these superfields on the mass shell.
In other words,
the classical super--p--branes propagate in a supergravity
background which satisfies supergravity equations of motion without
sources. Thus,
coupling of the complete D=11 supergravity action to the super--M--brane
worldvolume actions requires additional study. This problem does not arise
in the bosonic case, and in what follows we shall consider coupling of the
bosonic part of the duality--symmetric $D=11$ supergravity action to the
worldvolume actions of the membrane \ref\bst{E. Bergshoeff, E. Sezgin and
P. K. Townsend, {\sl Phys. Lett.} {\bf 189B} (1987) 75; {\sl Ann. Phys.
(N.Y.)} {\bf 185} (1988) 330.} and the five--brane \PSTU.

For this, we use an approach first
proposed by Dirac \DIR ~for describing electromagnetic interactions of
monopoles (see \ref\bs{
M. Blagojevi\`c and P. Senjanovi\`c, {\sl Phys. Rep.} {\bf 157} (1988)
233.} as a review), and further developed in \des \med.
The Dirac approach uses the fact that a charged
object (a particle, a string etc.) couples locally (and often minimally)
to the gauge field whose charge this object carries, and it couples (in
general) nonlocally to the dual gauge field strength by means of a
nonphysical Dirac string (or a p--brane) which stems from the charged
object.

Applying this method to the membrane we can couple it to both
the infinite field form and the nonpolynomial form of the
duality--symmetric $D=11$ supergravity action.

As to the five--brane, it
carries a two--form gauge field with a self--dual field strength in its
worldvolume, hence to construct an effective worldvolume action for the
five--brane one should apply one of the approaches discussed above. So far,
the covariant five--brane action has been constructed only in the
$a(x)$--field form \PSTU. We will see that this allows one to almost
straightforwardly couple the five--brane to the actions \six ~-- \cjs, but
not to \inf. For coupling to the latter, one should probably construct
an infinite field form of the five--brane action by generalizing
relevant results of \ref\ben{I. Bengtsson, {\sl Int. J. Mod. Phys.} {\bf
A12} (1997) 4869.} ~on a duality--symmetric formulation
of Dirac--Born--Infeld theory in $D=4$.

In the following subsections we shall discuss
features of ``$a(x)$--field'' coupling of $D=11$ supergravity and the
M--branes.

\subsec{Membrane coupling}
The worldvolume action for a membrane propagating in a curved
$D=11$ background and coupled to $A^{(3)}$ is \bst:
\eqn\acm{S_{{\cal M}_3}=-\int_{{\cal M}_3} d^3 y \sqrt{-det(g_{mn})}+
{1\over 2}\int_{{\cal M}_3} A^{(3)} \qquad  m,n = 0,1,2 }
where
\eqn\met{
g_{mn}(y)={{\partial x^{\up}}\over{\partial
y_m}}g_{{\underline{pq}}}(x(y))
{{\partial x^{\underline q}}\over{\partial y_n}}}
is an
induced worldvolume metric and $A^{(3)}(x(y))$ is the pullback of
$A^{(3)}(x)$ onto ${\cal M}_3$ (as in subsection 2.1).  For simplicity, we
have
put the membrane tension to one.

We can extend \acm ~to an integral over $D=11$ space--time by inserting
a $\delta$--function closed 8--form
\eqn\starj{
{~}^*J^3={1\over{3!8!}}
dx^{\um_1}\wedge...\wedge
dx^{\um_8}\epsilon_{\um_1...\um_8\un_1\un_2\un_3}\int_{{\cal
M}_3} d{\hat x}^{\un_1}\wedge d{\hat x}^{\un_2}\wedge d{\hat
x}^{\un_3}\delta(x-{\hat x}(y))
}
with the support on ${\cal M}_3$,
$J^{(3)\um\un\up}={1\over{\sqrt{-det(g_{\um\un})}}}\int_{{\cal M}_3}
d\hat x^{\um}\wedge d\hat x^{\un} \wedge d\hat x^{\up} \delta(x-{\hat
x}(y))$ being the membrane current minimally coupled to $A^{(3)}(x)$.

To couple the membrane action \acm ~to the duality--symmetric actions \six
~--\cjs ~we have to take care of the local symmetries \nine, \ten. These
are preserved if the field strength
$F^{(7)}=dA^{(6)}-A^{(3)}\wedge F^{(4)}$ in \six ~--\cjs ~ (except for the
Chern--Simons terms) is
extended to
\eqn\hatseven{\hat F^{(7)}=F^{(7)}-{~}^*G^{(4)}(x),}
where $G^{(4)}$ is defined by the equation
\eqn\dG{d{~}^*G^{(4)}={~}^*J^{(3)}.}
A solution to \dG ~is
\eqn\Gfour{G^{(4)\um_1...\um_4}(x)
={1\over{\sqrt{-det(g_{\um\un})}}}\int_{{\cal M}_4}d{\hat x}^{\um_1}\wedge
...\wedge d{\hat x}^{\um_4} \delta(x-{\hat x}(z)),}
where the integration
is performed over a four--dimensional surface ${\cal M}_4$ parametrized by
$z$, whose boundary is the membrane worldvolume ${\cal M}_3=\partial{\cal
M}_4$. This is the generalization of the Dirac string \DIR ~to a Dirac
three--brane stemmed from the membrane, by means of which the latter
couples to the dual gauge field strength $F^{(7)}$.

The complete action describing membrane coupling is therefore
\eqn\accomm{S=\int
d^{11}x{1\over 4}eR(\omega)~+~\hat S_A ~+~S_{{\cal M}_3},}
where $\hat S_A$  is either \six or \seven ~ with $\hat F^{(7)}$ \hatseven
~instead of $F^{(7)}$ everywhere except of the Chern--Simons term
$F^{(7)}\wedge F^{(4)}$.  If $\hat S_A$ is in the form \cjs, the minimal
coupling term in \acm ~doubles.

The equations of motion of $A^{(3)}$ and $A^{(6)}$ one gets from the
variation of \accomm ~reduce to the duality
conditions
\eqn\hatdc{\hat F^{(7)}=F^{(7)}-{~}^*G^{(4)}= {}^* F^{(4)}, \qquad
{~}^*\hat F^{(7)}=- F^{(4)}}
whose Bianchi identities are
the $D=11$ gauge field equations with the membrane source
\eqn\bianchi{d{}^* F^{(4)}+F^{(4)}\wedge F^{(4)}={~}^*J^{(3)},
\qquad  d{}^* \hat F^{(7)}=0.}
 From varying the metric, one gets
the Einstein equations:
\eqn\einst{R_{\um\un}-{1\over 2}g_{\um\un}R={1\over
3}(F^{(4)}_{\um\up_1\up_2\up_3}F_{\un}^{(4)\up_1\up_2\up_3} -{1\over 8}
g_{\um\un}F^{(4)}_{\up_1\up_2\up_3\up_4}F^{(4)\up_1\up_2\up_3\up_4})+
T^2_{\um\un},}
where the energy--momentum tensor of $A^{(3)}$ and $A^{(6)}$ reduces to
that of $A^{(3)}$ after taking into account the duality relation \hatdc,
and the last term on the right hand side of \einst ~is the membrane
energy--momentum tensor
$T^{2\um\un}=\sqrt{-det g_{pq}}g^{mn}\partial_mx^{\um}\partial_nx^{\un}$.
 From varying the membrane coordinate, one gets
the equation of motion of the membrane which is
\eqn\memeq{ {\partial\over {\partial y^m}}
(\sqrt{-det g_{pq}}g^{mn}(y)\partial_n x^{\ur}g_{\ur\uq})
-{1\over 2} g^{mn}\partial_mx^{\um}\partial_nx^{\un}
{\partial\over{\partial x^{\uq}}}g_{\um\un}(x)=}
$$
={1\over
2^.3!}\epsilon^{mnp} \partial_m x^{\um}\partial_n x^{\un}\partial_p
x^{\up}F^{(4)}_{\um\un\up\uq}.
$$

As expected, we have derived the standard equations of motion for
the bosonic fields of $D=11$ supergravity with the membrane as a source.

\subsec{Five--brane coupling}
The worldvolume action for the M--theory five--brane propagating in a
curved $D=11$ background and coupled to $A^{(3)}$ and $A^{(6)}$ is \PSTU
\BLN \schwarz:
\eqn\acbr{
S_{{\cal M}_6}=
\int_{{\cal M}_6} d^6 y \big[-\sqrt{-det(g_{mn} + i \tilde{H}_{mn})}
+\sqrt{-g}{1\over {4}}
v_l{~}^*H^{lmn}H_{mnp}v^p\big] }

$$ -
{1\over 2}\int_{{\cal M}_6}\left[A^{(6)}+ H^{(3)} \wedge
A^{(3)}\right], \qquad  m,n = 0,1,...,5 ,
$$
where $g_{mn}(y)$ is now the
induced metric of the five--brane worldvolume ${\cal M}_6$;
\eqn\sc{H^{(3)} =
dB^{(2)} -A^{(3)}= {1 \over 3!} dy^m \wedge dy^n \wedge dy^l H_{lnm}(y),
}
\eqn\h{H_{lmn}(y)= 3(\partial_{l}B_{mn}+
\partial_{m}B_{nl}+\partial_{n}B_{lm}) - A^{(3)}_{lmn} (x(y))
}
is the field
strength of the worldvolume gauge field $B_{mn}(y)$ which satisfies
the generalized self--duality condition on the mass shell:
\perry,
\eqn\tH{ \tilde H_{mn}\equiv v^l(y){~}^*H_{lmn}=(i_v{~}^*H^{(3)})_{mn},
\quad
{~}^*H^{mnl}={1\over{3!\sqrt{-g}}}\varepsilon^{mnlpqr}H_{pqr};
}
and
\eqn\five
{v_{m}(y)\equiv
{{\partial_{m}a(y)}\over{\sqrt{-(\partial^{l}a\partial_{l}a)^2}}},
\qquad v^{m}v_{m}=-1, }
where $a(y)\equiv a(x(y))$ is the pullback onto ${\cal M}_6$ of the
auxiliary scalar field $a(x)$ which we used in section 4 to construct the
covariant duality--symmetric $D=11$ supergravity action. In the 5--brane
action, the field $a(y)$ plays the same role \PSTU~ as in the
$D=11$ action. We should stress
that {\it a priori} one could try to use an independent worldvolume scalar
field to ensure the covariance of the five--brane action but, as we shall
see below, it turns out crucial for the consistent coupling of this action
to the duality--symmetric $D=11$ action \six ~-- \cjs ~that the auxiliary
worldvolume field is the pullback of $a(x)$.

The action \acbr ~is manifestly invariant under general coordinate
transformations of ${\cal M}_6$, and under the following local
transformations which transform $B^{(2)}$ and $a(y)$:
\eqn\brsymm{
\delta B^{(2)} =d\phi^{(1)}(y) +\phi^{(2)}(x(y))+da \wedge
\varphi^{(1)}(y) - {{\varphi(x(y))}\over{\sqrt{-(\partial a)^2}}} {\cal
H}^{(2)} , \qquad \delta a = \varphi(x(y)), }
where $\phi^{(1)}(y)$ is the
one--form parameter of the standard gauge transformations of $B^{(2)}$,
$\phi^{(2)}(x(y))$ is the pullback of the $D=11$ gauge transformations
\eight~ of $A^{(3)}$ which ensures that the field strength $H^{(3)}$ \sc
~is invariant under these transformations, $\varphi^{(1)}(y)$ parametrizes
a worldvolume analog of the transformations \nine ~and  ~implies an on
shell self--duality of $H^{(3)}$, $\varphi(x(y))$ is the pullback of the
corresponding $D=11$ scalar parameter of the local transformations \ten,
and \eqn\selfdB{ {\cal H}^{(2)} \equiv i_v H^{(3)} -
{{dy^n \wedge
 dy^m}\over\sqrt{-det g_{pq}}} ~{\delta \sqrt{det(g_{pq} + i
 \tilde{H}_{pq}) }\over{ \delta\tilde{H}_{mn}} }.}

 From \nine, \ten ~and \brsymm, we see that $D=11$ and ${\cal M}_6$ local
symmetries responsible for the duality properties of eleven--dimensional
supergravity and of the M--five--brane are intrinsically related to each
other.

The generalized non--linear self--duality condition on $H^{(3)}$,
\eqn\gsd{{\cal H}^{(2)} =0,}
is a consequence of the equation of motion of $B^{(2)}$ which follows
from \acbr  ~(see \perry, \PSTU ~for details).

We can extend \acbr ~to an integral over $D=11$ space--time by inserting
into it a $\delta$--function closed 5--form
\eqn\starji{
{~}^*J^6={1\over{5!6!}}
dx^{\um_1}\wedge...\wedge
dx^{\um_5}\epsilon_{\um_1...\um_5\un_1...\un_6}\int_{{\cal M}_6}d{\hat
x}^{\un_1}\wedge...\wedge{\hat
x}^{\un_6}\delta(x-{\hat x}(y))
}
with the support on ${\cal M}_6$,
$J^{(6)\uq_1...\uq_6}={1\over{{\sqrt{-det(g_{\um\un})}}}}\int_{{\cal
M}_6} d\hat x^{\uq_1}\wedge ...\wedge d\hat x^{\uq_6} \delta(x-{\hat
x}(y))$ being the five--brane current minimally coupled to $A^{(6)}(x)$.

To couple the five--brane action \acm ~to the duality--symmetric actions
\six ~, \seven ~we have to take care not only of the local symmetries
\nine, \ten, but also of \brsymm. Then the action describing consistent
coupling is
\eqn\accomfive{S=\int
d^{11}x{1\over 4}eR(\omega)~+~\hat S_A ~+~S_{{\cal M}_6}
-{1\over 2}\int H^{(3)}\wedge {\hat F}^{(4)}\wedge {~}^*G^{(7)}.}
In \accomfive, $\hat S_A$  is either \six ~or \seven ~where
everywhere, except in the Chern--Simons term, $F^{(4)}$ and $F^{(7)}$
are replaced with
\eqn\fhat{\hat F^{(4)}=F^{(4)}-{~}^*G^{(7)}, \quad
\hat F^{(7)}=F^{(7)}-H^{(3)}\wedge{~}^*G^{(7)},}
and the seven--form $G^{(7)}(x)$ is
defined by the equation
\eqn\dGs{d{~}^*G^{(7)}={~}^*J^{(6)}.}
A solution to \dG ~is
\eqn\Gseven{G^{(7)\um_1...\um_7}(x)=
{1\over{\sqrt{-det(g_{\um\un})}}}\int_{{\cal M}_7}d{\hat
x}^{\um_1}\wedge ...\wedge d{\hat x}^{\um_7} \delta(x-{\hat x}(z)),}
where
the integration is performed over a seven--dimensional surface ${\cal
M}_7$ parametrized by $z$, whose boundary is the five--brane worldvolume
${\cal M}_6=\partial{\cal M}_7$. This is the generalization of the Dirac
string \DIR ~to a Dirac six--brane stemmed from the five--brane, by
means of which the latter couples to the dual gauge field potentials
$F^{(4)}$ and $F^{(7)}$.

To check that the action \accomfive ~contains the symmetries
\eight ~-- \ten ~and \brsymm ~(where ~$\delta A^{(6)}$ acquires the
additional term $(da\wedge\varphi^{(1)}-{{\delta a}\over{\sqrt{-(\partial
a)^2}}} {\cal H}^{(2)})\wedge{~}^*G^{(7)}$),
we present the differential form of the
variation of \accomfive ~with respect to $A^{(3)}$, $A^{(6)}$, $B^{(2)}$
and $a$:
\eqn\varac{\delta S=
\int\big[\big(\delta A^{(6)}-\delta
A^{(3)}\wedge A^{(3)}- \delta B^{(2)}\wedge {~}^*G^{(7)}
-{{\delta a}\over{\sqrt{-(\partial a)^2}}}i_v
{\hat{\cal F}}^{(7)}\big)\wedge d(v\wedge i_v{\hat {\cal F}}^{(4)}) }
$$ +\big(\delta
A^{(3)}+{{\delta a}\over{\sqrt{-(\partial a)^2}}}i_v
{\hat{\cal F}}^{(4)}\big)\wedge
\big(d(v\wedge i_v{\hat{\cal F}}^{(7)})+2v\wedge i_v{\hat{\cal
F}}^{(4)}\wedge F^{(4)}
$$
$$ -v\wedge{\cal H}^{(2)}\wedge~{}^*J^{6}-
v\wedge i_v{\hat{\cal F}}^{(4)}\wedge{~}^*G^{(7)}\big)
$$
$$
-(\delta B^{(2)}+{{\delta a}\over{\sqrt{-(\partial a)^2}}}{\cal H}^{(2)})
\wedge \big(d(v \wedge {\cal H}^{(2)})
+ v \wedge i_v{\hat{\cal F}}^{(4)}\big)
 \wedge {~}^*J^{(6)}\big].
 $$
The equations of motion of $A^{(3)}$ and $A^{(6)}$ one gets from the
variation of \accomfive ~reduce to the duality
conditions
$$\hat F^{(7)}= {}^* \hat F^{(4)}, \qquad
{~}^*\hat F^{(7)}=-\hat F^{(4)}
$$
whose Bianchi identities are
the $D=11$ gauge field equations with the five--brane source
\eqn\bianchis{d{}^* \hat F^{(4)}+F^{(4)}\wedge \hat
F^{(4)}=H^{(3)}\wedge{~}^*J^{(6)},}
$$
d{}^* \hat F^{(7)}={~}^*J^{(6)}.
$$
The $D=11$ Einstein equations take the form:
\eqn\einste{R_{\um\un}-{1\over 2}g_{\um\un}R={1\over
3}(\hat F^{(4)}_{\um\up_1\up_2\up_3}\hat F_{\un}^{(4)\up_1\up_2\up_3}
-{1\over 8}
g_{\um\un}\hat
F^{(4)}_{\up_1\up_2\up_3\up_4}\hat F^{(4)\up_1\up_2\up_3\up_4})+
T^5_{\um\un},}
where  $T^{5\um\un}=
\sqrt{-det g_{pq}}T^{mn}\partial_mx^{\um}\partial_nx^{\un}$
is the energy--momentum tensor of the
5--brane.
An explicit form of a `formal' $d=6$ energy--momentum tensor $T^{mn}$ and
equations of motion of the 5--brane coordinate $x^{\um}(y)$ can be derived
from \ref\hsw{P.S. Howe and E. Sezgin {\sl Phys.  Lett.}
{\bf B394} (1997) 62\semi
P.S.  Howe, E.  Sezgin and P.  West, {\sl Phys.  Lett.}
{\bf B399} (1997) 49
\semi C. S. Chu and E. Sezgin, M--fivebrane from
the Open Supermembrane, hep-th/9710223.}
\ref\BAN{I.  Bandos, K. Lechner, A.
Nurmagambetov, P.  Pasti, D. Sorokin and M. Tonin,
{\sl Phys.Lett.} {\bf B408} (1997) 135.}.
We should note that the equation of motion \bianchis ~differs from an
analogous equation considered in \ALWIS ~ by a term ${~}^*G^{(7)}\wedge
{~}^*G^{(7)}$ which is absent in our version. Thus, eqs. \bianchis ~are
selfconsistent without the extra assumption of \ALWIS ~
that $d{~}^*G^{(7)}\wedge {~}^*G^{(7)}=0$. And, as we have already
mentioned, the equation of motion of $B^{(2)}$ produces the nonlinear
self--duality condition \gsd ~, while \ALWIS ~used its
linearized approximation (i.e.  $H^{(3)}={~}^*H^{(3)}$) which
was imposed by hand.

We can reduce the action \accomfive ~to an action which describes coupling
of the M--five--brane to the bosonic sector of the standard D=11
supergravity action. For this, we should first rewrite \accomfive
~in such a way that it will contain $\hat S_A$ in the form of \cjs
~with hatted $\hat F^{(4)}$  and $\hat F^{(7)}$ \fhat ~everywhere except in
the Chern--Simons term. This changes the form of five--brane
coupling, and the action takes the form
$$
S=\int d^{11}x~[{1\over 4}eR-
{e\over{2^.4!}}\hat F^{(4)}_{\um_1...\um_4}
\hat F^{(4)\um_1...\um_4}]
+\int {1\over 3}A^{(3)}\wedge dA^{(3)}\wedge dA^{(3)}
$$
\eqn\cjsc{ -\int d^{11}x~{e\over{2^. 3!}}v^{\up}{\hat{\cal
F}}^{(4)}_{\up\um_1\um_2\um_3}{\hat{\cal
F}}^{(4)\uq\um_1\um_2\um_3}v_{\uq}}
$$
-\int_{{\cal M}_6} d^6 y \big[\sqrt{-det(g_{mn} + i \tilde{H}_{mn})}
-{{\sqrt{-g}}\over {4}}
v_l{~}^*H^{lmn}H_{mnp}v^p\big]
$$
$$ -
{1\over 2}\int_{{\cal M}_6} dB^{(2)} \wedge
A^{(3)} + {1\over 2}\int A^{(3)}\wedge d A^{(3)}\wedge {}^*G^{(7)}.
$$
We see that the term of \acbr ~which described minimal coupling of the
five--brane to $A^{(6)}$ is replaced in ~\cjsc ~with a nonminimal
term $A^{(3)}\wedge d A^{(3)}\wedge {}^*G^{(7)}$, and $A^{(6)}$ remains
only inside of ${\hat{\cal F}}^{(4)}$. Now, imposing the gauge fixing
condition $i_v{\hat{\cal F}}^{(4)}=0$ (as in the free supergravity case of
subsection 4.2) we can eliminate the term with $A^{(6)}$ from \cjsc,
and the remaining action describes consistent nonminimal coupling of the
M--five--brane to the standard D=11 supergravity. This action reduces to
a corresponding version proposed in \ALWIS ~when the 5-brane action is
replaced with a quadratic term for $H^{(3)}$ \foot{ In the original
version of this paper, it was mistakenly claimed that the $A^{(3)}\wedge d
A^{(3)}\wedge {~}^*G^{(7)}$ term is absent from the action in \ALWIS.}.

Finally, coupling to D=11 supergravity of both a membrane and
a five--brane, with the membrane ending on the five--brane, is described
by the combination of actions  \accomm ~and \accomfive ~(or \cjsc) to
which (as in \ALWIS) one must add the membrane boundary term
\eqn\bound{\int_{\partial{\cal M}_3}B^{(2)}\equiv\int_{{\cal M}_6}
B^{(2)} \wedge {}^*j^{(2)}}
and extend the five--brane field strength
\sc ~to $\hat H^{(3)}=H^{(3)}-{}^*G^{(3)}$, where $d^*G^{(3)}={}^*j^{(2)}$
and $j^{(2)mn}(y^6)={1\over{\sqrt{-det g_{pq}(y^6)}}}\int_{{\cal
M}_2}d{\hat y}^{6 m}\wedge d{\hat y}^{6n} \delta(y^6-{\hat y^6}(y^2))$ with
$y^6$ and $y^2$ being, respectively, coordinates of ${\cal M}_6$
and ${\cal M}_2=\partial{\cal M}_3$.

\newsec{Conclusion} We have constructed a manifestly
duality--symmetric formulation of $D=11$ supergravity with the gauge
fields $A^{(3)}$ and $A^{(6)}$, coupled its bosonic sector to the
two--brane and the five--brane of M--theory, obtained corresponding
equations of motion of the system, and shown how the M--five--brane
couples to Cremmer--Julia--Scherk supergravity.  As further generalization,
one can consider coupling of $D=11$ supergravity to intersecting
M--branes.

We have found that consistent
coupling of the five--brane worldvolume effective action to the
supergravity action requires local symmetries responsible for duality
properties of the two actions to be
related to each other through the same auxiliary field.

The action for $D=11$ supergravity coupled to the M--branes might be
useful for a development of results
\ref\Witten{E. Witten, Five--brane
effective action in M--theory, hep--th/9610234.}
\ALWIS \ref\Bonora{L.
Bonora, C.  S.  Chu and M.  Rinaldi, Perturbative anomalies of the
M--5--brane, hep--th/9710063.} ~in studying anomalies in M--theory.

Methods developed in this paper can possibly
be applied to the study of coupling of
a self--dual IIB $D=10$ supergravity action \ref\dal{G. Dall'Agata, K.
Lechner and D. Sorokin, Covariant Actions for the Bosonic Sector of D=10
IIB Supergravity, hep-th/9707044.} ~to D--branes,
of their anomalies \ref\YIN{Y.--K. E. Cheung and Z. Yin, Anomalies,
Branes and Currents, hep--th/9710206.}
 ~and vacuum energy \ref\sodin{T. Kadoyoshi, S. Nojiri, S. D. Odintsov and
A. Sugamoto, Vacuum polarization of supersymmetric D-brane in the
constant electromagnetic field, hep--th/9710010.}, in particular, in the
case of a self--dual D--3--brane \ref\tse{A. Tseytlin, {\sl Nucl. Phys.}
{\bf B469} (1996) 51.}~  which admits a manifest duality--symmetric
description \ben \ref\ber{D. Berman, SL(2,Z) duality of Born--Infeld
theory from non--linear self--dual electrodynamics in 6 dimensions,
hep-th/9706208.}, as well as to the consideration of analogous problems in
Type IIA superstring theory.

We should note that in this paper we dealt with classical systems.
As has been stressed by Witten
\Witten,
one should expect to find a problem
when quantizing any action
for a self-dual system. For example, for a $d=2$ chiral scalar,
the path integral can not be defined in
a modular-invariant manner, implying that the amplitude must depend
on more than just the metric of the two-dimensional surface.
In the first approach to constructing actions,
this problem probably comes from
subtleties involving the infinite number of fields. In the second approach,
the problem probably comes from trying to define the correct phase
space for the a(x) - field \ref\wp.
{E. Witten, private communication.}. In any case, this quantization
problem is relevant for the actions proposed in this paper and
certainly deserves further study.

\bigskip
\noindent
{\bf Acknowledgements}. The authors would like to thank K. Lechner,
P. Pasti, S. Randjbar--Daemi, and M. Tonin for fruitful
discussions, H. Nishino for pointing out an error in the supersymmetry
transformations, E. Witten for pointing out difficulties with quantizing
self-dual actions,
S. de Alwis for comments, and
P. Townsend for suggesting this work.
I.B.  and D.S.  acknowledge partial support from grants of the
Ministry of Science and Technology of Ukraine and the INTAS Grants N
93--493--ext, N 94--2317 and N 96--0308.  N.B. acknowledges partial
support from CNPq grant number 300256/94.

\newsec{Appendix: SUSY transformation of $A^{(6)}$}

To justify that the supersymmetry variation of $A^{(6)}$ has the right
form,
let us extend the field strengths \scf ~and \scs ~to
a curved $D=11$ superspace $Z^M=(x^{\um}, \Theta^\mu)$ where, to describe
supergeometry, one introduces supervielbeins
$E^A(Z)=dZ^ME_M^{~A}=(E^{\ua},E^\alpha)$. Note that $E^{\ua}|_{\Theta=0}
=e^{\ua}$
and $E^\alpha|_{\Theta=0}=\Psi^\alpha$.

The analysis of Bianchi identities \DAF,
\ref\CL{A. Candiello and K. Lechner, {\sl Nucl. Phys.} {\bf B412} (1994)
479.  }
shows that superfield strengths, which at $\Theta=0$ reduce to
\scf ~and \scs, have only vector supervielbein components
\eqn\scff{ \tilde
F^{(4)}={1\over {4!}}E^{\ua} E^{\ub} E^{\uc} E^{\ud} \tilde
F^{(4)}_{\ud\uc\ub\ua}\equiv dA^{(3)}-{1\over 2}E^{\ua}  E^{\ub} E^\a  E^\b
(\Gamma_{\ua\ub})_{\a\b}, }
\eqn\scfs
{\tilde F^{(7)}= {1\over {7!}}E^{\ua_1} \dots
E^{\ua_7} \tilde F^{(7)}_{\ua_7\dots \ua_1} }
$$\equiv dA^{(6)}-A^{(3)}\wedge
dA^{(3)}- {2\over {5!}}E^{\ua_1}\dots E^{\ua_5}  E^\a  E^\b
(\Gamma_{\ua_1\dots \ua_5})_{\a\b},$$
where the last terms in \scff ~and \scfs
~replace, respectively, $C^{(4)}$ and $C^{(7)}$ of \scf ~and \scs ~with
their superform counterparts.

Now, by definition, a
p--rank superform $W^{(p)}$ varies under supersymmetry transformations as
follows:
\eqn\sv
{\delta_\epsilon W^{(p)}=i_\epsilon dW^{(p)}+d(i_\epsilon W^{(p)}),}
where $i_\epsilon$ defines the contraction of the spinor supervielbein
components of the superform with the supersymmetry parameter
$\epsilon^\a$. Hence, because of the constraints \scff ~and \scfs, the
second term in \sv ~is absent from the supersymmetry variations of
$\tilde F^{(4)}$ and $\tilde F^{(7)}$ (this just implies that they are
supercovariant).
Using these properties and substituting
superfield analogues of the
transformations \susy~ and \sus~
into the right hand side of the
supersymmetry variation of \scfs,
one can convince oneself that they
correctly reproduce the supersymmetry variation \sv ~of the left hand
side of \scfs.

\listrefs
\end